\RequirePackage[2020-02-02]{latexrelease}
\documentclass[twocolumn,prl,aps]{revtex4}
\usepackage{amsmath}
\usepackage{amssymb}
\usepackage{graphicx}
\usepackage{dcolumn}
\usepackage{bm}
\usepackage{xcolor}
\usepackage{epstopdf}

\begin{document}
\title{Thermoelectric transport in a three-dimensional HgTe topological insulator}

\author{G. M. Gusev, $^1$  Z. D. Kvon, $^{2,3}$  A. D. Levin, $^{1}$ and  N. N. Mikhailov $^{2,3}$}

\affiliation{$^1$Instituto de F\'{\i}sica da Universidade de S\~ao
Paulo, 135960-170, S\~ao Paulo, SP, Brazil}
\affiliation{$^2$Institute of Semiconductor Physics, Novosibirsk
630090, Russia}
\affiliation{$^3$Novosibirsk State University, Novosibirsk 630090,
Russia}

\date{\today}
\begin{abstract}

The thermoelectric response of 80-nm-thick strained HgTe films of a three-dimensional topological insulator (3D TI) has been studied experimentally. An ambipolar thermopower
 is observed where the Fermi energy moves from conducting to the valence bulk band. The comparison
between theory and experiment shows that the thermopower is mostly due to the phonon drag contribution.
In the region where the 2D Dirac electrons coexist with bulk  hole states, the Seebeck coefficient is  modified due to 2D electron - 3D hole scattering.
\end{abstract}

\maketitle
\section{Introduction}

A three dimensional topological insulator (3D TI) has gapless surface
state inside the bulk band-gap \cite{qi, qi2, hasan1, hasan2}. The surface
state energy spectrum has the form of a Dirac cone, which holds massless particles.
Remarkably the spin of surface Dirac electrons is locked perpendicular to the wave vector $k$ in the 2D plane,
which leads to the suppression of the  electron scattering on impurities.
The wide strain HgTe films are among of the best host 3DTI materials  \cite{brune, brune2}  because, in such a system, a very
high mobility of 2D surface electrons $\mu\sim100 m^{2}/Vs$ is achieved \cite{kvon1, kozlov1, kozlov2, kozlov3}.

\begin{figure}[ht!]
\includegraphics[width=\linewidth]{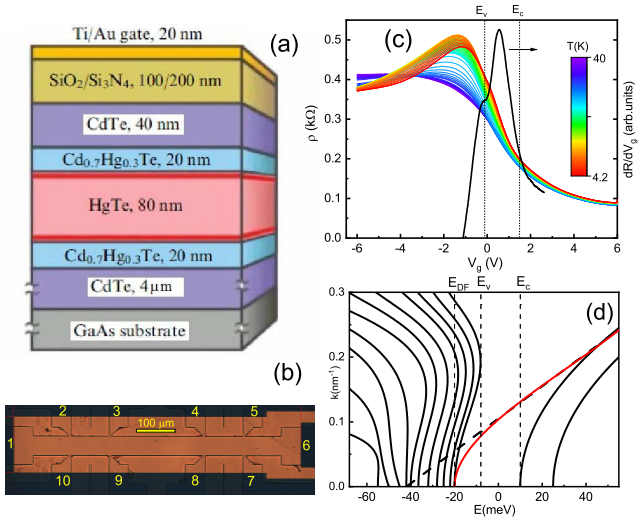}
\caption{(Color online)(a) Schematic of the transistor and (b) Top view of the sample. (c) Resistivity $\rho$  as a function of gate voltage measured for different temperatures.
The derivative of the resistance $dR/dV_{g}$ as a function of gate voltage at T=4.2K.
(d) Schematic of the energy spectrum of a strained 80 nm mercury telluride film. Conduction
and valence band edges are marked by $E_{c}$ and $E_{v}$, the edge band of the surface states (the Dirac point), which is located in the valence band, by $E_{DF}$. Dashes represent the spectrum of interface states
 under approximation where mixing of these states with bulk hole states is neglected.}
\end{figure}

The thermoelectric measurements can probe the sign of the charge carriers and the transport mechanisms
and are widely used to obtain complementary information about electron transport in metals and semiconductors.
Moreover, the value of the thermoelectric coefficient strongly depends on the energy spectrum and the mechanism of the time relaxation.
For example, an important relationship exists between the diffusive thermopower $S_{xx}$ and the logarithmic derivative of the
longitudinal electric conductivity $\sigma_{xx}$ of a metal:
\begin{equation}
S_{xx}= -\frac{\pi^{2}}{3e}k_{B}T\frac{d}{d\mu}\left[\ln \sigma_{xx}(\mu)\right]
\end{equation}

where $\mu$  is the chemical potential of charge carriers. Based on the Mott relation, anomalously large
thermopower has been predicted in a 2D topological insulator \cite{xu}. A 2D TI
is characterized by a pair of counterpropagating gapless edge modes inside of the bulk gap\cite{konig}. These edge
states have helical spin properties and are proposed to be robust to backscattering \cite{qi, hasan1, gusev1}.
It is suggested \cite{xu} that,  when  the  Fermi  level  approaches  the  conduction or valence band
edge, the scattering rate of electrons in the helical one dimensional modes increases significantly due to 1D-2D scattering, which leads to
anomalous growth of the amplitude of the Seebeck signal and a change of its sign. Note, however, that this mechanism
requires complete suppression of the scattering between the edge states, which is not observed in realistic structures \cite{gusev1}.
An experimental probe of the thermoelectric response demonstrates that the observed thermopower is mostly due to the bulk contribution, while the resistance is determined by both the
edge and bulk transport \cite{gusev2}.

A similar situation is expected in a 3D TI: when the Fermi level crosses the edge of the bulk bands, additional 2D-3D scattering can lead to an increase in
thermopower coefficients. Such mutual scattering has been detected directly in the resistance behaviour \cite{kozlov1, kozlov2}.
Note, that recently 2D electron - 3D hole scattering has been deduced from nonmonotonic differential
resistance of narrow 3DTI HgTe channels  \cite{muller}.

Another system, where the coexistence of the two distinct types of carriers with a different charge sign affects the transport properties, is
a 2D semimetalic HgTe well of intermediate well width ($\sim 20 nm$) \cite{kvon, olshanetsky}. In this system, 2D electron - 2D hole scattering  directly results in temperature dependent resistivity $\rho$, which increases with temperature as $\rho\sim T^{2}$ in accordance with the prediction for electron-hole friction coefficient behaviour \cite{entin}.
The thermopower in such a system has been studied in papers \cite{gusev3, olshanetsky2}.  A comparison between theory and experiment demonstrated
that the observed thermopower in a 2D electron-hole system is mostly due to phonon drag. It has been argued that the role of 2D electron-2D
hole scattering is important in the formation of  thermoelectric power mechanisms.

Thus, thermoelectric power is a very important tool to study the mechanism of scattering between carriers of different signs  and even between carriers
of different dimensions, as for example 1D-2D ( 2D topological insulators \cite{xu})and  2D electron- 2D holes ( 2D semimetals \cite {gusev3}).

In the present paper, we report an experimental study of the thermoelectric response in 80 nm thick strained HgTe layers.
We found that thermopower in a 3DTI  is due to phonon drag, which is similar to a 2D semimetal system in 20 nm HgTe wells. When the Fermi level
crosses the region with coexistence of 2D electrons and 3D holes, mutual scattering causes a strong change in thermopower.

\section{Materials and Methods}
The HgTe layers were grown by molecular beam epitaxy on (013)-oriented (GaAs). The sample was a 80-nm HgTe layer that is
sandwiched between two $Hg_{0.3}Cd_{0.7}$ Te buffer layers above and below
(40 nm) (figure 1a). The details of the structural properties of the prepared sample has been published in previous paper \cite{dvoretsky}.
For transport measurements, a field effect transistor was used. The
sample  was  a  long  Hall  bar  consisting  of  three  50  $\mu m$
wide  consecutive  segments  of  different  length  (100,  250,
and 100 $\mu m$) and 8 voltage probes (figure 1b).
The top of the Hall bar device was covered by a dielectric layer and subsequently a  metallic gate.
A pyrolytic $SiO_{2}$ layer or a double $SiO_{2}  + Si_{3}N_{4}$ layer grown at temperatures of $80-100^{\circ} C$ was
used as a dielectric, and the TiAu layer served as a gate.
The resistance measurements
were performed in a variable temperature insert
cryostat in the temperature range 1.4–50 K using the
standard  four  point  scheme.  The
electrically powered heater was glued symmetrically
near  Contact  1  (see  figure 2 left panel)  and created a
temperature  gradient  in  the  system,  while  the  other
end  was  indium  soldered  to  a  small  copper  slab  that
served  as  a  thermal  ground.  One  calibrated  thermo  sensor  was  attached  at
the end of the sample near the heater while the other
was attached to the heat sink (see  figure 2 left panel ). Thermo sensors were
used to measure the $\Delta T$ along the sample. For typical heat power, we applied $\nabla T=160 K/m$. The voltages
induced by this gradient were measured by a lock-in
detector  at  the double  frequency  of  $2f_{0}=0.8-2  Hz $ across
various  voltage  probes. The thermovoltage was proportional to the applied power. Thermal  conductance  of
the sample was overwhelmingly dominated by phonon
transport in the GaAs substrate \cite{gusev2}. 3 samples were measured.

\section{Results}

Figure 1c presents the  resistances at zero
magnetic field for the sample fabricated from a 80 nm HgTe layer for different temperatures.
The current I flows between Contacts 1 and 6, and the voltage V is measured between the short distance
separated Probes 2 and 3, $R = R_{1,6}^{2,3}$ (Figure 1b). The temperature dependence $R(V_{g},T)$ reveals
the different character of the transport in the different regions of the energy spectrum, and has been
performed in the previous publications \cite{kozlov1, kozlov2}. Fig 1d shows the specific features in the energy spectrum of the strained HgTe layer.
It is worth noting that the lattice constant of  CdTe is  slightly larger than that of HgTe, inducing strain on the sample.
The strain opens an energy gap in the energy spectrum. The Dirac point in the 3D TI ($E_{DF}$) is
located  deep in the valence band and due to hybridization with the valence band, the spectrum of the
surface states, consisting of Dirac electrons near the bottom of the valence band,
deviates from the linear law (Fig 1d, red line). Thus, when the gate voltage is sweeping from
negative to positive values, the electrochemical potential $\mu$ moves from the conductance band ($\mu >E_{c}$),
through the bulk gap ( $E_{v}< \mu < E_{c}$ ) to the valence band ($\mu < E_{v}$), as one can see
in Figure 1d. Note, however, that the energy scale in Figure 1d does  not necessarily
correspond with the gate voltage scale in Figure 1c  due to nonmonotonic behaviour of Fermi energy with density. For example the point
$E_{DF}$ corresponds to the gate voltage $V_{g}=-20 V$ because of the high density of the states in the valence band.
Since dielectrics break down may occur at lower gate voltages, we can not approach this energy point.
Resistance in the bulk gap region originates from the helical surface electron states,
with different densities in the top and the bottom surface \cite{kozlov1}. In the region $E _{DF}< \mu < E_{v}$,
two dimensional surface electrons and 3D bulk holes coexist. Nonmonotonic temperature dependence of the resistance
is observed: R(T) increases for temperatures below 15 K, while decreasing above 20 K. We attribute this behaviour to
2D electron-3D holes scattering, which is similar to 2D electron-2D hole scattering in HgTe semimetal wells \cite{olshanetsky, entin}.
Figure 1d shows that the derivative of resistance $dR/dV_{g}$ reveals the features in points $E_{c}$ and $E_{v}$, correspondingly.
In the region $\mu < E_{DF}$, the transport is determined by 3D bulk holes, and we don't expect peculiarities in
resistance and thermopower behaviour. The position of the energy $E_{DF}$ can be obtained approximately  from a detailed analysis of the Shubnikov de Haas oscillations \cite{kvon1}.

\begin{figure}[ht!]
\includegraphics[width=\linewidth]{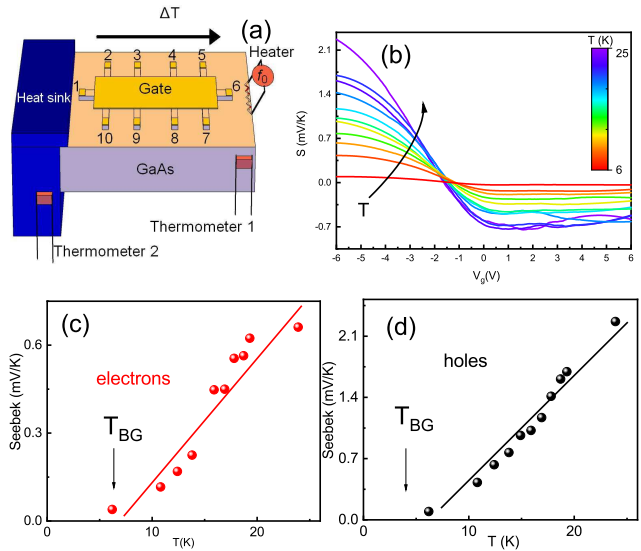}
\caption{(Color online) (a) Sample geometry. (b) Seebeck coefficient as a function of the gate voltage for different temperatures. (c) Temperature dependence of Seebeck coefficient at $V_{g}=6 V$ (electrons).
(d) Temperature dependence of Seebeck coefficient at $V_{g}=-6 V$ (holes). The solid lines correspond to $S_{xx}\sim T$. Arrows indicate Bloch-Gruneisen temperature.}
\end{figure}

Once we have determined the gate voltage and the density interval with different transport properties, we now turn to
study the thermoelectric response in our films. Experimentally measured
quantities were the longitudinal thermovoltage  $V_{xx}=S_{xx}\nabla T L$ where $L=450 \mu m$
 is the distance between Probes 1 and 6 along the temperature gradient $ \nabla T$ (figure 2a).
 We also measured longitudinal thermovoltage between Probes 2-3 and 2-5 in order to
 examine the homogeneity of the temperature gradient, and found reasonable proportionality to L.
 Particularly we found the ratio of the signal $V_{1,6}/V_{2,3} =6-8$, and $V_{1,6}/V_{2,3} =1.5-2$ which approximately agrees with the distance between probes.

\begin{figure}[ht!]
\includegraphics[width=\linewidth]{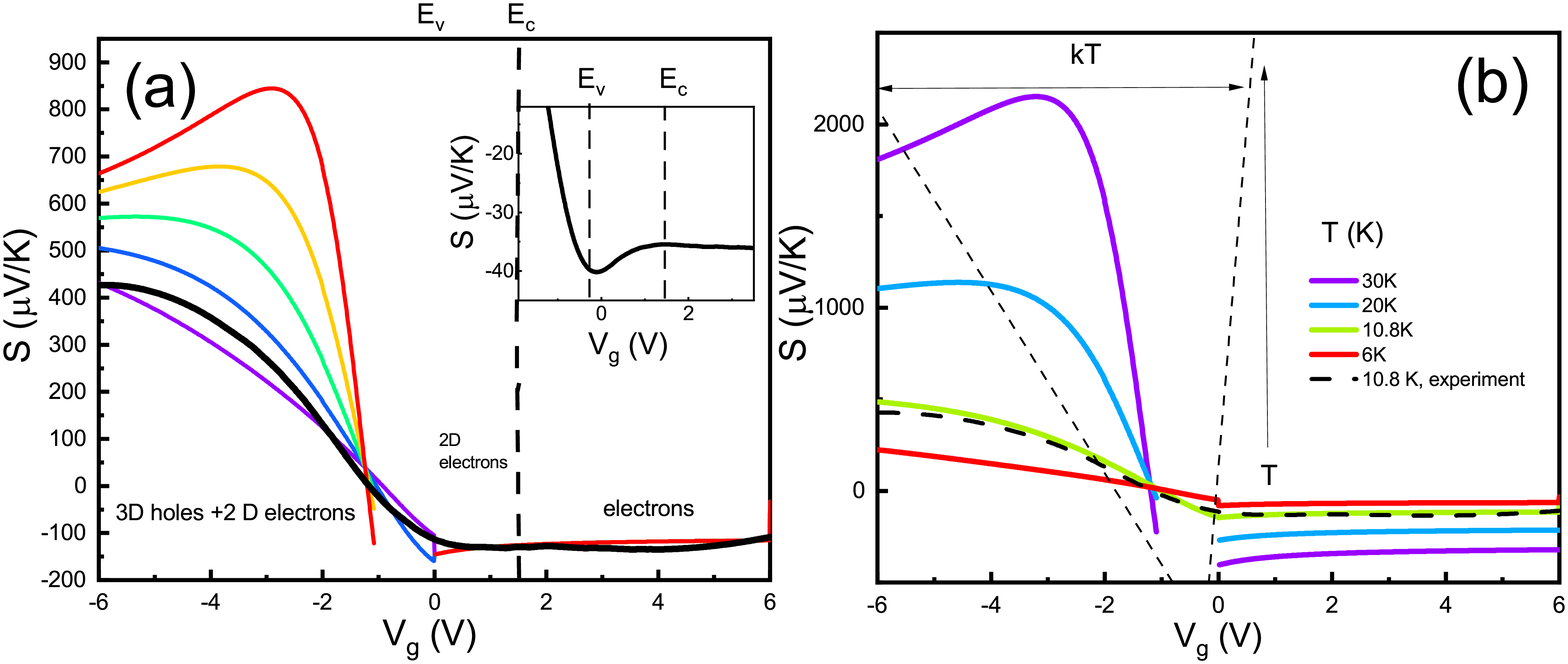}
\caption{(Color online)(a) Seebeck coefficient as a function of the gate voltage calculated for eqs. 2-8 with parameters indicated in the text and for different parameter $\Theta$: 0.1, 0.25, 0.5, 1, 2, 4 ($\times4.8\times10^{-38}
 Joule*s/K^{2}$). Black line is the Seebeck coefficient measured at T=10.8 K. Insert shows $V_{g}$ dependence of the Seebeck coefficient zoomed-in on the voltage interval $E _{v}< \mu < E_{c}$ at T=4.2 K. (b) Seebeck coefficient as a function of the gate voltage calculated for eqs. 2-8 with parameters indicated in the text and for different temperatures. Horizontal arrows shows the interval where $\mu\sim kT$, and where degenerate approximation for electrons and holes is not valid more. Therefore the equations (2-8) can not  be applied to this region. Dashed line corresponds to the measured Seebeck coefficient at 10.8 K. }
\end{figure}

Figure 2b shows the gate  voltage  dependencies  of  the  Seebeck coefficient $S_{xx}$ at  different  temperatures in zero magnetic field.
The thermopower is negative in the region $E_{v}< \mu $. A further decrease of the gate voltage towards hole contribution, Seebeck coefficient changes the sign
crossing the zero at the voltage corresponding to the transition from electron dominant to hole dominant contribution, which is not coincident neither with
charge neutrality point, determined from zero Hall resistance measurement in a magnetic field (\cite{kozlov2, kozlov3}), neither with  positions of energies $E_{v}$ or $E_{DF}$.
The value of the Seebeck coefficient is larger for the holes. The figure 2b  displays the traces of $S_{xx}$ versus  $V_{g}$  for  different  temperatures.
Figure 2c,d   shows the temperature dependence of the Seebeck coefficient module  measured at selected gate voltages  $V_{g}$=6 and -6 V, corresponding to the
electron and hole dominated regions. One can see that  $S_{xx}$ grows linearly with temperature as $S_{xx} \sim T$  in the interval $10 < T < 25 K$.
In the region $E _{DF}< \mu < E_{v}$, where 2D electrons and 3D bulk holes coexist, we are not able to distinguish the
thermopower mechanism due to electron-hole scattering, which can modify the temperature dependence of $S_{xx}(T)$, and for which a comparison between the theory and the experiment
requires more advanced theory. We should note that previous research in 3DTI does not reveal semimetallic behaviour and features in the resistivity and thermopower due to e-h frictions,
since samples have a lower mobility \cite{josta}.

\section{Discussions}
Detailed calculations of thermopower in 2D semimetals  have been performed in papers \cite{gusev3, olshanetsky2}, where both contributions ( diffusion and phonon drag)
have been taken into account. The Seebeck coefficient in zero magnetic field is given by:

\begin{equation}
S_{x x}=\frac{\Lambda_{x}}{Z}
\end{equation}

\begin{equation}
\begin{aligned}
Z=&\left[m_{e} n_{h} \tau_{h}+m_{h} n_{e} \tau_{e}+\left(n_{e}-n_{h}\right)^{2} \eta \tau_{e} \tau_{h}\right]^{2} \\
&
\end{aligned}
\end{equation}

\begin{equation}
\begin{gathered}
\Lambda_{x}=\frac{1}{e}\left\{\left[A^{e} \tau_{e} m_{h}-A^{h} \tau_{h} m_{e}+\left(A^{e}+A^{h}\right)\left(n_{e}-n_{h}\right) \eta \tau_{e} \tau_{h}\right]\right. \\
\times\left[m_{h} \tau_{e} n_{e}+m_{e} \tau_{h} n_{h}+\left(n_{e}-n_{h}\right)^{2} \eta \tau_{e} \tau_{h}\right]\left.\right\}
\end{gathered}
\end{equation}

where  $A^{e,h}=A_{\mathrm{dif}}^{e,h}+A_{\mathrm{ph}-\mathrm{dr}}^{e,h}$ are the electron and hole terms corresponding to the diffusion and phonon-drag contributions to the
Seebeck coefficient respectively, $n_{e}$ and $n_{h}$ are the
electron and hole densities, $m_{e} = 0.03m_{0}$ and $m_{h} = 0.3m_{0}$—the electron and hole effective mass, $g_{e}$ = 1 and $g_{h}$ = 2—the electron
and hole valley degeneracy, $\tau_{e}$ and  $\tau_{h}$ —the electron and hole transport scattering time, determined from the electron and
hole mobilities respectively, $k$—Boltzmann constant, $\eta = \Theta\times T^{2}$—is the electron-hole friction coefficient. Parameter $\Theta$ is introduced in
paper \cite{olshanetsky}  for scattering between 2D electrons and 2D holes. It is determined by  the peculiarities of the interaction between electrons and holes and dependents of their densities.

The diffusion contribution does not contain any adjustable parameters  and is given by
\begin{equation}
A_{\mathrm{dif}}^{e,h}=-\frac{\pi}{3 \hbar^{2}} k^{2} T m_{e,h} g_{e,h}
\end{equation}

The phonon drag contribution depends on the material specific
phonon relaxation rate and the temperature regime.  The system enters into
the Bloch-Gruneisen (BG) regime at a very low temperature T when acoustic
phonon wave vector $q =2k_{F}$, where $k_{F}$ is the Fermi wave vector.
In our HgTe system, we found that for the densities $n_{s}\sim 10^{12} cm^{-2}$ the characteristic temperature
$T_{BG}=\frac{2k_{F}s\hbar}{k}$ ( $s$ is the sound velocity) is around 4-6 K ( figure 2c,d).  If we propose  that $\tau_{ph} = const$, then for
temperatures $ T \gg T_{BG}$, we obtain:

\begin{equation}
A_{\mathrm{ph}-\mathrm{dr}}^{e,h}=-\frac{1}{3 \hbar^{5}} k \tau_{\mathrm{ph}} m_{e,h}^{2} s p_{F _{e,h}}^{2} g_{e,h} B_{e,h}\left(q_{T}\right)
\end{equation}
where $q_{T} = kT /\hbar s$ is the thermal phonon wave vector and for acoustic
phonons in cubic crystals the function $B_{e,h}(q_{T})$ reads:
\begin{equation}
B_{e,h}(q_{T})=\frac{\Lambda_{e,h}^{2} q_{T}}{2 \delta s}
\end{equation}

here $\Lambda_{e,h}$ are the deformation potential constants, $\delta$ is the crystal
density. The data for the deformation potential:
$\Lambda_{e}=-4.8 eV$ and $\Lambda_{h}=-0.92 eV$ for $\delta$ and $s$ we have:
$\delta=8.2 g/cm^{3}$, $s=3.2\times10^{5} cm/s$. One can see that in the temperature regime $T\gg T_{BG}$ both contributions $A_{\mathrm{dif}}^{e,h}$ and $A_{\mathrm{ph}-\mathrm{dr}}^{e,h}$
linearly depend on temperature.

Note that in the monopolar limit in the regions $\mu < E_{DF}$, where the transport is determined by 3D bulk holes, we obtain the following
equations for diffusive and  drag contributions (using subscript $h$ ):
\begin{equation}
S^{e,h}_{diff, ph-dr}=\pm\frac{A_{\mathrm{dif, ph-dr}}^{e,h}}{e n_{e,h}}
\end{equation}

In the conduction band in the region $\mu > E_{c}$, the bulk and surface carriers coexist, however, for simplicity
we applied equation (8)(using subscript $n$ ), where $n_{e}$ is the total bulk and surface densities. For simplicity, we consider the situation when
the top and bottom surfaces are equally occupied. Since our HgTe layer is not thick, bulk carriers can be considered as a quasi-two dimensional system (see
Figure 1d).

Figure 3a shows the comparison between the theoretical  Seebeck coefficients calculated according to eqs.(2-8) for different parameter $\Theta$  and the experimental curve measured at T=6 K as a function of $V_{g}$.
Parameter  $\Theta$ is associated with the electron-hole friction coefficient $\eta$, and is responsible for features in thermopower in the region $E_{DF}< \mu < E_{c}$, where
surface electrons and bulk holes coexist. Figure 3b shows the comparison between the theoretical  Seebeck coefficients calculated according to eqs.(2-8) for different temperatures as a function of $V_{g}$.
We obtain good quantitative and qualitative agreement with the theory.

The diffusion contributions in the monopolar regions $S^{e,h}_{diff}$ were calculated according to Eq.(5) for the sample parameters, determined from conductivity. No adjustable parameters have been used in the calculation
of the diffusive thermo emf. The phonon drag contribution in the monopolar regions $S_{e,h}^{ph-dr}$  was calculated according to Eqs. (6)–(8), using a constant phonon relaxation time $\tau_{ph}=0.6\times10^{-7} s$, corresponding to the relaxation length $l_{ph}= s \tau_{ph}=0.2 mm$, which is close to the sample size. We argue  that phonon relaxation length is determined by their scattering on the substrate boundaries. Since the diffusion contribution was calculated without adjustable parameters, we found that $S^{e,h}_{diff}<< S^{e,h}_{ph-dr}$, therefore, one may conclude here that  the phonon drag contribution is the dominant contribution at high temperature $T>4.2K$. As we already mentioned above, the temperature dependence of Seebeck coefficient is linear for temperature above $T_{BG}$ for both contributions.

In the bipolar region $E_{DF}< \mu < E_{c}$, where surface electrons and bulk holes coexist, thermopower was calculated according to eqs.(2-7). We used a simplified model, assuming that the electron-hole friction coefficient $\eta$
does not depend on the electron and hole densities. Indeed this model is much too simple to adequately describe the shape of thermopower behaviour in this region. Figure  3a demonstrates that
the shape of the curve $S (V_{g})$ is closer to the experimental one for $\Theta=1.0\times10^{-38} Joule s/K^{2}$, which is smaller than friction coefficient for 2D electron and 2D hole system
$\Theta=4.8\times10^{-38} Joule s/K^{2}$ \cite{entin}.  The insert shows Seebeck coefficient zoomed-in on the voltage interval $E _{v}< \mu < E_{c}$ for T=4.2K.
One can see that thermopower is enhanced near point $\mu \simeq E_{c}$, which we attribute to the 2D Dirac electron -3D bulk holes scattering.  The feature near $E_{v}$ is smear out  at higher temperatures.
Note, that the equations in the monopolar regime (8) cannot be obtained from Equation (3,4) from the transition to the monopolar case.
It is because they are obtained under the assumption that Fermi  gases  are  degenerate.  Indeed,  the  transition  to
the  monopolar  limit  at  low  temperatures  occurs  in  a relatively narrow range of the chemical potential $\Delta \mu \sim T$ ( see figure 3b). It may lead to discontinuity in the calculated thermopower
around transition points.  While our experiment offers an interesting outlook on
thermopower in this region, more experimental and theoretical work is required to understand the behavior of the friction between 2D electron and 3D holes in a 3D topological insulator.

It is worth noting, that we also compared the experiment with the monopolar model (eqs. 8),  considering independent  2D electron and 3D hole contribution to thermopower only (not shown here).
Indeed we obtained considerable disagreement between the theory and the experiment, which supports the evidence  of mutual electron- hole friction in our system \cite{kvon1, kozlov1, muller}.

Figure 3b shows the temperature dependence of the theoretical curves  $S (V_{g})$. As we expected,  in the  monopolar region, $S (V_{g})$ is proportional to the temperature in accordance with
experimental observations (Figure 2c,d), while in the bipolar region $E_{DF}< \mu < E_{c}$, the Seebeck coefficient $S (V_{g})$  grows with temperature faster due to mutual friction temperature dependence $\eta = \Theta\times T^{2}$.
We don't see such behaviour in the experiment Fig.2b. As we mentioned above, the model is valid for degenerate Fermi gases, and cannot
be applied at high temperatures near charge neutrality points.  More advanced theory is required to describe this behaviour, which is out of the scope of our experimental paper.

\section{Conclusions}

In conclusion, this work is the first to study the behavior of the
thermo emf of a  three dimensional topological insulator based on an HgTe 80nm thick film.
The obtained experimental dependencies are compared with a theory.
When the gate voltage is sweeping from
negative to positive values, the electrochemical potential $\mu$ moves from the conductance band ($\mu >E_{c}$),
through the bulk gap ( $E_{v}< \mu < E_{c}$ ) to the valence band ($\mu < E_{v}$),
and we expect a different thermopower regime. In the monopolar regimes,
we demonstrate the calculated values of the transport coefficients corresponding to the drag contribution, which is
approximately an order of magnitude larger than diffusion thermopower. Taking this contribution into account,
we determine the phonon relaxation length, which
turns out to be temperature independent and caused by phonon
scattering at the structure boundaries.
In the bipolar region $E_{DF}< \mu < E_{c}$, where surface electrons and bulk holes coexist, Seebeck coefficient is  modified due to 2D electron - 3D hole scattering.
Comparison with the theory demonstrates good agreement, however, exact knowledge of the mutual friction behaviour is required for better understanding
of thermopower in such a nontrivial regime.

\section{Acknowledgment}
The financial support of this work by the Ministry of Science and Higher Education of the Russian Federation, Grant No. 075-15-2020-
797 (13.1902.21.0024);  São Paulo Research Foundation (FAPESP) Grant No. 2015/16191-5, and the National Council for Scientific and Technological Development (CNPq) is acknowledged.
We  thank  E.B.Olshanetsly, M. V. Entin, L. I. Magarill and O.E. Raichev  for  the  helpful  discussions.

\end{document}